\documentclass[preprint,showpacs,preprintnumbers,amsmath,amssymb]{revtex4}

\usepackage{latexsym,amssymb,makeidx}
\usepackage{epsfig, color, ulem}
\usepackage{graphicx}
\usepackage{amsmath}

\begin{document}

\begin{center}
\rule{0cm}{2cm}
{\bf \LARGE A Geothermal Well Doublet for Research and Heat Supply of the TU Delft Campus}\\
\rule{0cm}{1cm}\\
Philip J. Vardon, David F. Bruhn, Abe Steiginga, Barbara Cox, Hemmo Abels, Auke Barnhoorn, Guy Drijkoningen, Evert Slob, Kees Wapenaar
\end{center}

\hspace{0cm}\textbf{Summary}\\

A geothermal well doublet, designed for two primary aims of research and commercial heat supply, is planned to be constructed on the campus of Delft University of Technology. The plans include a comprehensive research programme, including installation of a wide range of instruments alongside a comprehensive logging and coring programme and an extensive surface monitoring network. The wells will be cored, with samples from all representative geological units down to the reservoir. An extensive suite of well-logs is planned to provide detailed information on the properties of the various units. Fibre-optic cables will be installed in both wells all the way down to the reservoir section, which is anticipated to be at 2200m, in the Lower Cretaceous Delft Sandstone. The Delft sandstone is well-known as a reservoir rock for natural gas in the West Netherlands Basin. The wells will be operated by a commercial entity, but the infrastructure is designed and explicitly installed as a research infrastructure. As such it will become part of the European EPOS (European Plate Observing System, https://www.epos-ip.org), such that accessibility and data availability will not be limited to TU Delft researchers. The university has made a decision in principle to start drilling no later than 2020.

\pagebreak

\section{INTRODUCTION}

Though the number of geothermal well doublets in operation has been increasing over the last years, none of these wells has been equipped with more than the bare minimum of monitoring equipment, and for none of them were detailed geological and geophysical base line data collected. Despite the success of implementing this technology at a reasonable scale, many questions related to practical and more fundamental operational aspects are therefore still open. These include classical issues such as the uncertainty about productivity and temperature of the resource prior to drilling, but also problems related to downhole processes and fluid chemistry and operational decision making. These fundamental issues often lead to operational problems such as poor injectivity, corrosion of installations and precipitation of solids. In addition, our limited understanding of the thermal, hydraulic, mechanical and chemical processes in the subsurface related to geothermal operations result in sub-optimal exploitation of the resource and potentially in technically and/or economically failed projects. Moreover, the link to heat demand and heat distribution in a heterogeneous system (such as an urban environment) has not been extensively studied.

Worldwide, research on deep geothermal wells has been focused on high-enthalpy heat production (water $>$ 100°C; primarily aimed at electricity production), developments in the Netherlands have targeted low-enthalpy geothermal aquifers (i.e. water $<$ 100°C) for domestic and agricultural heating purposes. A geothermal doublet has been proposed to be constructed on the TU Delft campus (e.g., Bruhn et al. 2015), known as the DAPwell. It is designed to provide sustainable heat source for the TU Delft campus buildings and as a living laboratory facility for state-of-the-art science and education.

Drilling is planned to take place in 2020. In addition, the heating systems of existing TU Delft buildings need to be retrofitted to allow for lower temperature input than is currently the case. In a first stage, only buildings on campus will be supplied with heat from the well. If there is a clear demand and wish to connect other buildings on/near the campus to the heating system, the project might be expanded at a later stage.

The DAPwell doublet will be a unique subsurface research infrastructure planned to work in this energy range and will therefore make a significant contribution to close the knowledge gap concerning specific questions related to low-enthalpy geothermal energy production. It will provide the opportunity to develop and test technical solutions in a realistic, urban environment.

\section{RESEARCH PROGRAMME}

The research programme has received funding to provide the research infrastructure via the EPOS-NL project. The research activities associated with this are made up from a series of ongoing and future projects, such that an overall programme can evolve over the lifetime of the well to answer contemporary salient questions.

\subsection{EPOS-NL}

The project EPOS-NL is the Dutch contribution to the European Plate Observing System (EPOS; www.epos-ip.org), which is a European organization facilitating the integrated use of data, data products and infrastructure for Earth science. EPOS-NL integrates groups and facilities from Utrecht University (UU) and Delft University of Technology (TUD) with the Royal Netherlands Meteorological Institute (KNMI), which is the national centre for seismological and acoustic data services. The objective of EPOS-NL is to implement infrastructure for frontier-breaking research on geo-resources and geo-hazards, in particular on geo(thermal)-energy, geological storage of energy and CO2 and earthquakes induced from human activities. EPOS-NL will comprise a cluster of facilities including:
\begin{itemize}
\item	A Groningen gas field seismological network and data centre (KNMI).
\item	The Earth Simulation Laboratory for multi-scale, rock physics and analogue experiments (UU).
\item	A distributed facility for multi-scale imaging and tomography (MINT) of geo-materials.
\item	The campus deep geothermal research facility (TUD).
\end{itemize}
Incorporation of EPOS-NL within the European EPOS research infrastructure facilitates transnational access to physical facilities, as well as optimal exploitation of research results via open access data services.

\subsection{Main questions / ambition}

With the DAPwell research programme we will create a world-class geothermal energy research facility where the following key topics will be addressed:
\begin{enumerate}
\item		Predictive power of flow models for optimal control and monitoring.
\item		Hydraulic-Thermal-Chemical behaviour:
\begin{itemize}
\item		Chemistry of geothermal fluids and their interaction with reservoir rocks and technical installations.
\item		Monitoring of travelling fluid and cold fronts.
\item		Advanced inflow estimation.
\end{itemize}
\item		Effect of human activities in the subsurface on the natural and built environment at surface.  
\item		Novel well completion: composite casing material.
\item		Integrated domestic heat management. 
\item		Site specific characterisation: geological history, heterogeneities, reservoir fluids (water/gas/oil).
\end{enumerate}
Research undertaken in conjunction with the DAPwell will result in knowledge for further innovation in geothermal science and engineering on topics that, at the moment, hamper the ability to fully exploit the potential of this technology. 

The research will target a better prediction of the lifetime of a geothermal doublet, i.e. when the breakthrough of the cold front at the production well occurs, meaning that heat production substantially reduces. The approach suggested is (a) geophysical monitoring using the full range of active-source electromagnetics, passive electromagnetics and seismics, chemical tracers, cross-well pressure transient analysis, with (b) data-assimilation methods to integrate the results into prediction tools. In addition, the system integrity will be improved, with an innovative approach to prevent corrosion. The approach proposed is to (a) investigate none-corrosive casing materials and study their long-term performance, (b) understanding the fluid chemistry as a function of pressure and temperature, and (c) developing modelling tools to predict the chemical interaction of the complex fluid with rocks and casing. Moreover, the wider energy system will be investigated. The designed doublet will exceed the heat demand of the campus and provide heat when there is no demand, therefore the management and extension of the heat grid to outside the campus will be investigated. 

In addition, we anticipate that new scientific questions and need for innovations will occur during the lifetime of this large-scale long-term project. Therefore, flexibility is built into the scientific design.

\section{PROJECT DESIGN}

The DAPwell is currently at the final conceptual design stage, prior to detailed design. Three partners are working together in the project, TU Delft, EBN and Hydreco Geomec. For more details on the project progress and commercial details, please see www.tudelft.nl/geothermalwell. 

The Delft Sandstone is the target formation. This formation is well known as geothermal reservoir in the south-western part of the Netherlands. The nearby Pijnacker doublet has been producing from this formation since 2010. Within a radius of 5km from the Delft Geothermal wellsite there are a total of 36 wells or sidetracks. Most of the data from these wells are publicly available on www.nlog.nl. Recently drilled Pijnacker geothermal (PNA-GT) wells are located in the adjacent concessions; not all data from these wells are publicly available yet, but available through the cooperation between the DAP project and the PNA-GT projects.

\begin{figure}[h]
\vspace{-1cm}
\centerline{\epsfysize=10. cm \epsfbox{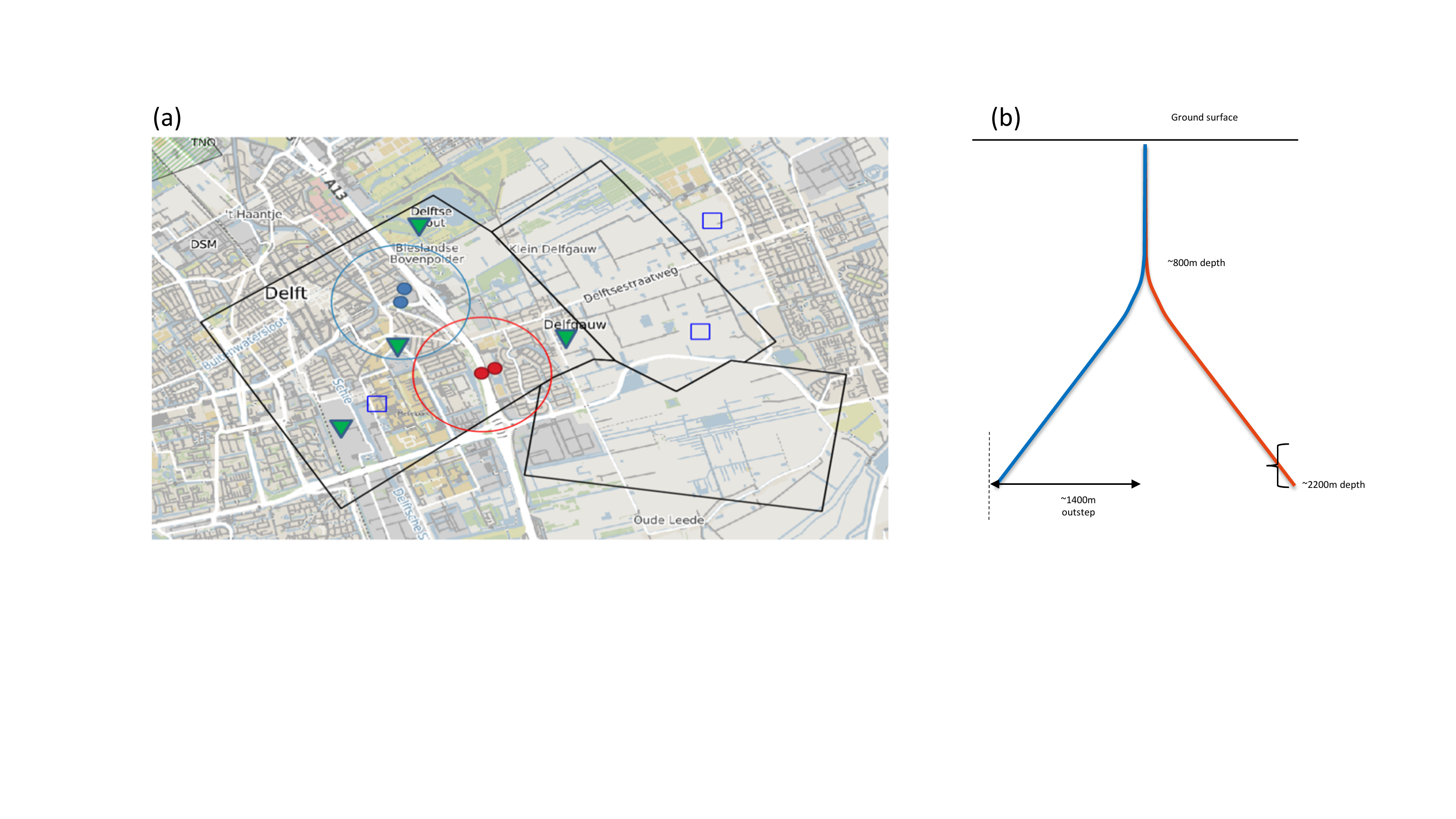}}
\vspace{-3cm}
\caption{\small (a) The designed geothermal location, with the reservoir top and bottom indicated by circles (red for producer and blue for injector). The black boxes are the license areas, including two adjacent projects and the green triangles are the proposed surface geophysical shallow borehole array locations. (b) Designed well trajectory, note that the outsteps are not on the same plane. 
}\label{Fig1}
\end{figure}

The doublet is designed to have a flow rate of approximately 320m$^3$/hr and produce hot water at approximately 75°C. Current planning foresees that the return temperature to the system will be initially around 50°C, although it is anticipated that this will reduce over time, as buildings are renovated and further buildings are included in the heat distribution system. TU Delft will withdraw at least 100GJ of heat energy per year to provide heating to the campus. To enable an easy connection to the existing heat grid, the surface location of the wells is planned to be close to the existing heat grid in a location where sufficient space is available for the required construction infrastructure, indicated approximately in Figure 1(a) by the open blue square on the left hand side. The wells will follow the schematic trajectory shown in Figure 1(b), where vertical sections from the ground surface to approximately 800 m depth followed by deviated sections at approximately 45° until the reservoir. This means that each well will outstep by approximately 1400 m, both towards where the reservoir is deeper (i.e. not in the opposite direction from each other), as indicated in Figure 1(a) by the red and blue circles, indicating the well intersection with the top and bottom of the reservoir. A 3D representation of the reservoir is shown in Figure 2.

\begin{figure}[h]
\centerline{\epsfysize=8. cm \epsfbox{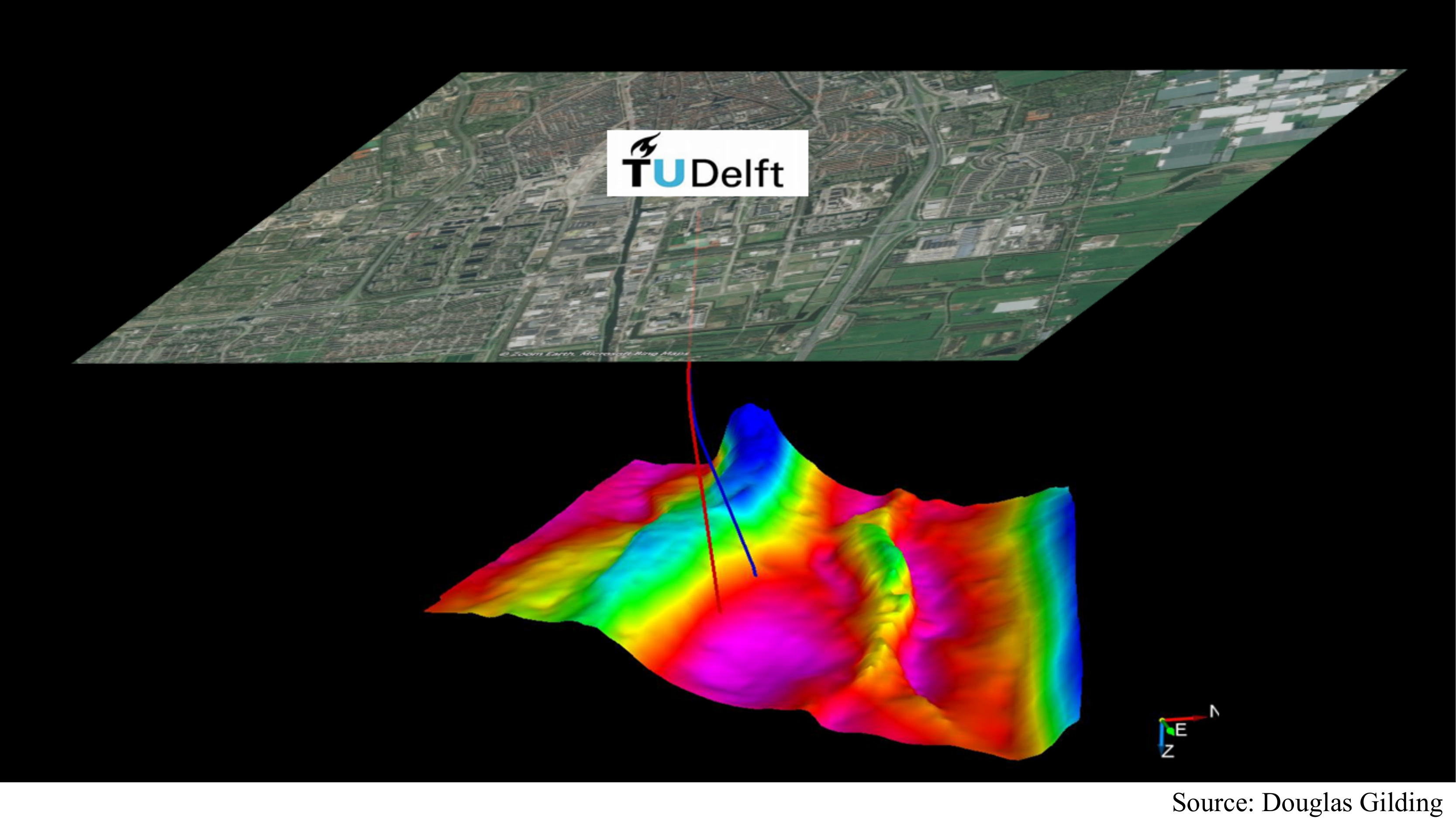}}
\caption{\small 3D view of the reservoir including the planned wells and existing geothermal projects and exploratory wells. 
}\label{Fig2}
\end{figure}

The diameter of the production well is planned to be approximately 8$\frac{5}{8}$ inch in the reservoir section, with a glass reinforced epoxy (GRE) lined steel tubing foreseen above the reservoir in the lower straight (deviated) section, increasing in size until the surface. An external casing is planned, which will result in an annulus between the production tubing and external casing. This annulus will be used for (i) monitoring to detect the performance of the casing and production tubing and (ii) as a carrier for the fibre optic monitoring system. This means that the fibre optic could be replaced if needed or if new instrumentation is proposed. A submersible pump is designed to be inserted at approximately 850 m depth with a production tubing to surface. The injection well is proposed to be approximately 9$\frac{5}{8}$ inch in the reservoir with a liner running to surface. Larger sections (liners) will be constructed closed to surface. This well is also proposed to be GRE lined from above the reservoir to the surface (WEP, 2019). 

The geology derived from seismic lines and exploration wells nearby, is shown in Figure 3. The proposed coring plans are shown alongside. Due to the interest in oil and gas exploration, there is less known in the region between approximately 200 – 1700 m depth. In particular there is known to be a series of unconformities at around 400 m deep. The upper layers are also of interest for heat storage. The complete reservoir, along with the cap rock and underburden are planned to be cored. 

\begin{figure}[h]
\centerline{\epsfysize=8. cm \epsfbox{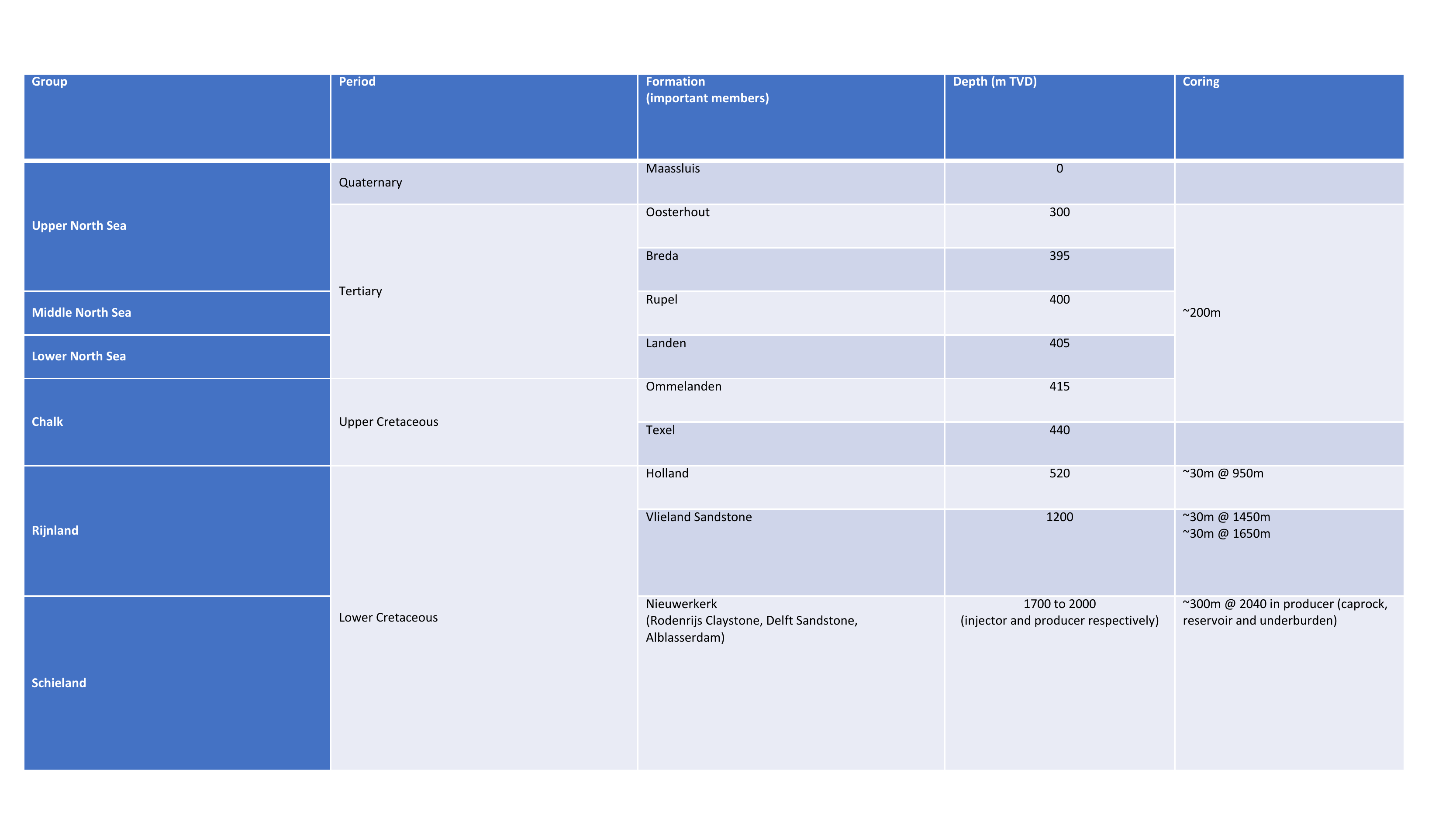}}
\vspace{-1cm}
\caption{\small Simplified lithography (after WEP, 2014) and coring plans.  
}\label{Fig4}
\end{figure}

\section{LOGGING AND MONITORING PROGRAMME}

One of the main objectives of the research is to fully understand the production and behaviour of the flow in the reservoir. To do this well logging – both prior to casing and afterwards – and ongoing monitoring is planned in both wells.

The proposed logging programme in the open holes will include
\begin{itemize}
\item	SP and natural gamma-ray (plus spectral gamma-ray)
\item	Resistivity 
\begin{itemize}
\item	deep, shallow and micro laterolog (LLD, LLS, microL)
\item	Induction log - IL 
\item	micro spherically focussed log – MSFL
\item	dual-spacing neutron log (CNL) with compensated formation density log (FDC)
\end{itemize}
\item	Sonic ($v_p$, $v_s$, full waveforms)
\item	Gamma Density
\item	Nuclear magnetic resonance logging – NMR (maybe, for permeability)
\item	Micro dip meter tool, dip measurements 
\item	Fullbore micro-imager logs - FMI
\item	Mini-frac tests at certain levels 
\item	Caliper log
\end{itemize}

Inside the casing: 
\begin{itemize}
\item	Gamma ray 
\item	Induction logs 
\item	FDC / CNL 
\item	Sonic 
\item	Ultrasonic Imaging Tool (USIT) and casing collar locator (CCL) 
\item	Nuclear magnetic resonance logging – NMR (maybe)
\item	A cement bond log (CBL) to test the cementation of the novel casing, and an accurate caliper
\end{itemize}

The programme may still have to be adapted based on risk and objectives. Some of the above tools can be combined in one toolstring to reduce the openhole time. Fluid samples will be taken to perform pressure, volume, temperature (PVT) analysis of the reservoir fluids, after the first production test. Samples will have to be taken by means of wireline with memory gauge or E-line. For this the ESP will have to be removed after the production test. 

Standard production logging will take place, alongside a bespoke monitoring system. The monitoring system is mainly based upon fibre optic technology. In the injection well, the fibre optic cable (a bundle of fibre optics) will be installed behind the inner liner, cemented in place to ensure a solid contact with the formation. This cable is designed to measure distributed temperatures and acoustics. The acoustic sensors are planned to be wound in a way such that multidirectional acoustics can be detected. In the production well, the fibre optic cable will be attached to the outside of the production casing, which finishes above the reservoir, and will be supported inside the well through the reservoir. This cable will enable distributed temperature and acoustic sensing (DTS and DAS), and additionally will include distributed pressure sensing within the reservoir section. The pressure sensors will allow (i) temporal pressure analysis and (ii) differential flow measurement. These measurements will be augmented by occasional downhole flow monitoring and local tracer measurements. 

To monitor the impact at the surface and downhole two vertical fibre optic cables will be installed, and a dense seismic and electromagnetic monitoring network will be set up, providing a new basis for observation of processes and changes occurring following the operation of deep geothermal heating doublets. The design is intended to detect any induced seismicity and allow the source to be spatially identified. Temperature and induced flow (convection) monitoring is planned in the shallow subsurface. Innovative methods of seismic interferometry and virtual seismology under continuous development (e.g., Boullenger et al. 2015, Wapenaar et al. 2018) will be applied to determine the location, source mechanism and radiation characteristics of seismicity to image and monitor cold-water injection in DAPWELL. The proposed locations for the geophysical array are shown by green triangles in Figure 1(a). 

The borehole seismic measurements using the fibre-optic cables will be carried out in time-lapse mode using an active source, and in continuous monitoring mode using the acoustic noise generated by the water production/injection. As active seismic source, a highly innovative seismic vibrator based on linear synchronous motors will be used that has been developed at TU Delft (Fig. 4a). This source will be placed at the surface and allows generation of low-distortion, very repeatable and wide-frequency-band signals, including the low frequencies.

\begin{figure}[h]
\centerline{\epsfysize=10. cm \epsfbox{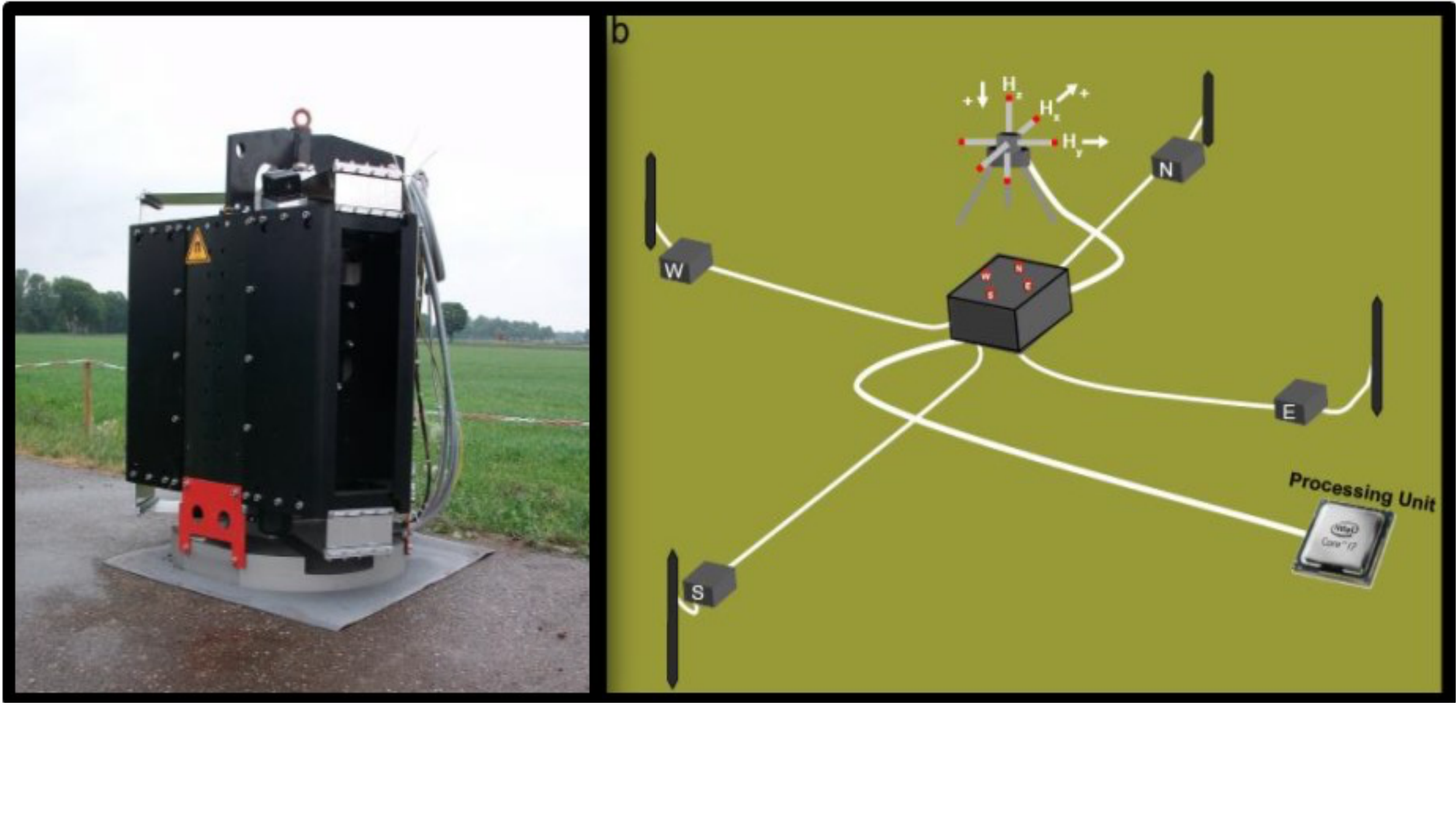}}
\vspace{-1cm}
\caption{\small a) Innovative seismic vibrator based on linear synchronous motors that generates low frequencies (typically down to 1.5 or 2 Hz); b) Sketch of the electromagnetic network. 
}\label{Fig3}
\end{figure}

With active-source electromagnetics, high-resistivity bodies can be detected, localized, and monitored from the surface and with sensors in the wells (Wirianto et al., 2011, Schaller et al., 2019). Producing large volumes of hot and injecting cold water creates acoustic and electromagnetic noise, which can be used as signal in so-called passive (ambient-noise) measurements. Using advanced signal processing methodology developed at TU Delft, these passive measurements can be used in addition to the active-source electromagnetic measurements to improve the imaging and monitoring around the well. The major challenges are the detection of a body that is relatively small compared to its depth and the influence of anthropogenic noise in an urban environment.

For the active electromagnetic (EM) monitoring a transient electric-dipole source will be used at the surface, while electric and magnetic sensors will be placed on the surface (Fig. 4b) and in the boreholes. Pseudo random binary sequences (PRBS) will be used as the source time-signal designed to improve the signal-to-noise ratio. This equipment can be bought with possible adaptations to the source time signature. Downhole electrodes for electric field measurements will be developed and placed for passive monitoring using electromagnetic noise generated by the water production/injection. 

The development of electrodes and housing of receiver electronics to endure permanent exposure to the ambient high pressure and temperature and the aggressive chemical composition of the pore fluids is an outstanding technical challenge. No sensors seem to exist at the moment that can operate continuously for long periods under these conditions. In case the development of sensors that stand these conditions fails, alternative borehole sensors will be made that can be operated in time-lapse mode and will be removed from the borehole after the transient measurements. Electrodes are mostly sensitive to corrosion and the challenge is to find a solution that provides resistance to corrosion while maintaining low contact resistance over time. Receiver electronics housing will be designed and built in house with existing expertise. The most crucial development concerns the realisation of a stable electrical contact between the electrodes and the terminals of the housing.

\section{LABORATORY INFRASTRUCTURE}

To make sure the wells can serve as reference wells for research, an extensive coring programme will be implemented during drilling. The cores will be stored and made available for laboratory testing, general characterisation and calibration of subsurface measurements. The petrophysical properties of the reservoir rocks will be determined in the Petrophysics Laboratory at TU Delft, such as density, porosity and permeability, electrical and thermal conductivities as well as mechanical properties such as elastic moduli. The detailed mineralogical and chemical analysis (X-ray methods, microscopy, SEM and electron microprobe, in EPOS-NL MINT) will provide the basis for the evaluation of fluid-rock interaction and for the monitoring and prediction of the long-term behaviour of the system. 

\section{CONCLUSIONS}

The DAPWELL research and monitoring infrastructure will be used to investigate the fundamental scientific challenges that are presently limiting the development of geothermal energy. This could not be realized by using an existing geothermal plant that is built for commercial energy production because the monitoring equipment cannot be installed once operation has started and drill cores are not taken in commercial wells. This exceptional development will result in innovations that leapfrog from well-functioning doublets used today to highly efficient geothermal installation in ten years’ time. 

Access to the high-level research infrastructure will render DAPWELL in particular and EPOS-NL in general not only a key national infrastructure but also result in major appeal to talented researchers from elsewhere in Europe and beyond. Likewise, the university partners participate in a wide range of European programmes that will assure visibility, which is prerequisite for high-potential researchers to become interested in EPOS-NL and to be attracted by geothermal research. 

The unique aspect of the DAPWELL facility will be the possibility to do research using an operating geothermal plant. It will be used as a laboratory where researchers study optimal production scenarios and monitoring techniques in order to achieve the highest possible energy efficiency. This could not be realized by using an existing geothermal plant that is built for commercial energy production. The unique approach of research at DAPWELL will result in innovations to improve the efficiency of geothermal systems. As there is currently no such research infrastructure in an operating geothermal well doublet worldwide, the DAPWELL will provide a unique opportunity for many high-potential researchers to do research with this fully operational research well. The uniqueness of the infrastructure will lead to productive cooperation, especially with European partner institutions (for example, the EERA joint programme) and the growing geothermal industry in the Netherlands and beyond.

\section*{ACKNOWLEDGEMENTS}

We acknowledge funding from The Netherlands Organization
for Scientific Research (NWO), National Roadmap Programme ``EPOS-NL: The Netherlands contribution to the European Plate Observing System''.
In addition, the research of K. Wapenaar has received funding from the European Research Council (ERC) under the European Union’s Horizon 2020 research and innovation program (grant no. 742703).

\section*{REFERENCES}

\begin{enumerate}
\item Boullenger, B., Verdel, A., Paap, B., Thorbecke, J., Draganov, D.: Studying CO2 storage with ambient-noise seismic interferometry: A combined numerical feasibility study and field-data example for Ketzin, Germany, Geophysics, Vol. 80 (1), Q1-Q13 (2015).
\item Bruhn, D.F., Wolf, K.-H., Woning, M., van Dalen, E., Nick, H.M., Willems, C.J.L., Hellinga, C.: The Delft Aardwarmte Project (DAP): Providing Renewable Heat for the University Campus and a Research Base for the Geothermal Community, Proceedings World Geothermal Congress 2015, Melbourne, Australia, 19-25April 2015.
\item Schaller, A., R. Streich, G. Drijkoningen, O. Ritter, and E. Slob, 2018, A land-based controlled-source electromagnetic method for oil field exploration: An example from the Schoonebeek oil field, Geophysics,83(2), WB1-WB17.
\item Wapenaar, K., Brackenhoff, J., Thorbecke, J., van der Neut, J., Slob, E., Verschuur, E.: Virtual acoustics in inhomogeneous media with single-sided access, Scientific Reports, Vol. 8, 2497 (2018).
\item Wirianto, M., W. A. Mulder and E. C. Slob, 2010, A feasibility study of land CSEM reservoir monitoring in a complex 3D model, Geophysical Journal International, 181, 741-755.
\item WEP: Detailed Well Construction Design (DAP-GT-01 \& DAP-GT-02), Technical Report (2014).
\item WEP: Concept Well Design: Delft Aardwarmte Doublet, Technical Report, version 1.6 (2019).
\end{enumerate}

\end{document}